\documentclass{ecai}

\usepackage{graphicx}
\usepackage{latexsym}
\usepackage{hyperref}
\usepackage{pgfplots}
\pgfplotsset{compat=1.3}
\usepackage{pgfplotstable}
\usepackage{tikzscale}
\usepackage{xcolor}
\usepackage[caption=false]{subfig}
\usepackage{flushend}
\usepackage{tikz}
\usetikzlibrary{shapes,arrows,positioning}
\usetikzlibrary{calc}

\newcommand{\SOMA}[0]{SOMA}

\definecolor{barMainCol}{RGB}{0, 48, 82}
\definecolor{barCSLfood}{RGB}{112, 184, 93}
\definecolor{barCSLdishes}{RGB}{40, 125, 120}
\definecolor{barIAIfood}{RGB}{28, 155, 216}
\definecolor{barIAIdishes}{RGB}{0, 100, 163}
\definecolor{situationBack}{RGB}{84, 208, 255}
\definecolor{actionBack}{RGB}{180, 230, 250}
\definecolor{motionBack}{RGB}{220, 245, 255}
\definecolor{accentred}{RGB}{231, 76, 60}
\definecolor{accentyellow}{RGB}{241, 230, 100}
\definecolor{accentorange}{RGB}{253, 143, 0}
\definecolor{accentgreen}{RGB}{112, 184, 93}
\definecolor{lightblue}{RGB}{28, 155, 216}
\definecolor{darkblue}{RGB}{0, 48, 82}
\definecolor{easeblue}{HTML}{144F78}
\definecolor{outer}{HTML}{0d2f54}
\definecolor{inner}{HTML}{bee4fb}
\definecolor{light}{HTML}{e2f3fe}

\def\horzBarXlim{30}
\def\horzBarWidth{6}
\def\horzBarY{7}

\newsavebox\MyPicture
\NewDocumentCommand{\roundedpicture}%
      {O{width=0.6\linewidth}
       O{draw=white!80!easeblue,line width=0.5pt,rounded corners=8pt}
       m}{%
   \savebox\MyPicture{\includegraphics[#1]{#3}}%
   \begin{tikzpicture}%
    \draw [path picture={%
                   \node[anchor=center] at (path picture bounding box.center) {%
                       \usebox\MyPicture};},#2]
          (0,0)  rectangle (\wd\MyPicture,\ht\MyPicture);
   \end{tikzpicture}%
}

\begin{document}
\begin{frontmatter}

\title{Towards a Neuronally Consistent Ontology for Robotic Agents}

\author[A]{\fnms{Florian}~\snm{Ahrens}\thanks{Corresponding Author. Email: fahrens@uni-bremen.de}}
\author[B]{\fnms{Mihai}~\snm{Pomarlan}}
\author[C]{\fnms{Daniel}~\snm{Be{\ss}ler}}
\author[A]{\fnms{Thorsten}~\snm{Fehr}}
\author[C]{\fnms{Michael}~\snm{Beetz}}
\author[A]{\fnms{Manfred}~\snm{Herrmann}}

\address[A]{University of Bremen, Department of Neuropsychology
and Behavioral Neurobiology}
\address[B]{University of Bremen, Institute for Linguistics}
\address[C]{University of Bremen, Institute for Artificial Intelligence}

\begin{abstract}
The Collaborative Research Center for Everyday Activity Science \& Engineering (CRC EASE) aims to enable robots to perform environmental interaction tasks with close to human capacity.
It therefore employs a shared ontology to model the activity of both kinds of agents, empowering robots to learn from human experiences.
To properly describe these human experiences, the ontology will strongly benefit from incorporating characteristics of neuronal information processing which are not accessible from a behavioral perspective alone.
We, therefore, propose the analysis of human neuroimaging data for evaluation and validation of concepts and events defined in the ontology model underlying most of the CRC projects.
In an exploratory analysis, we employed an Independent Component Analysis (ICA) on functional Magnetic Resonance Imaging (fMRI) data from participants who were presented with the same complex video stimuli of activities as robotic and human agents in different environments and contexts. We then correlated the activity patterns of brain networks represented by derived components with timings of annotated event categories as defined by the ontology model.
The present results demonstrate a subset of common networks with stable correlations and specificity towards particular event classes and groups, associated with environmental and contextual factors. These neuronal characteristics will open up avenues for adapting the ontology model to be more consistent with human information processing.
\end{abstract}
\end{frontmatter}

\section{Introduction}
The development of autonomous robotic agents by the Collaborative Research Center for \emph{Everyday Activity Science \& Engineering} (CRC EASE) is based on the principles of cognition enabled robotic control, employing systems for self-reflected reasoning and planning~\cite{beetz_cognition-enabled_2012, beetz_know_2018}.
Subsystems of the project’s cognitive architecture thereby interact with a central knowledge base which is populated not only by the robots’ own experiences but also by recorded environmental interactions of humans in the real-world as well as virtual reality contexts~\cite{beetz_open-ease_2015, beetz_know_2018}.
The goal of this effort is to build a knowledge base from experience -- whether simulated, observed, or self-performed.

Data in the knowledge base are stored as 
\emph{Narrative-Enabled Episodic Memories} (NEEMs) which are subdivided into experience and narrative. The NEEM narrative provides a symbolic description of a NEEM in terms of its semantic nature, e.g., the action of picking up a cup, while the NEEM experience contributes the corresponding multimodal sub-symbolic data of the acting agent -- human or robot -- in the form of, amongst others, audio- and video-recording, motion vectors, captured peripheral physiological parameters or brain activity derived signals.
Through this linkage of sub-symbolic and symbolic domains, the robot will be enabled to query task specific experiences, allowing it to adjust the way how an abstract task is executed based on previous experiences of successful executions.

\begin{figure}[t]
    \centering
    \input{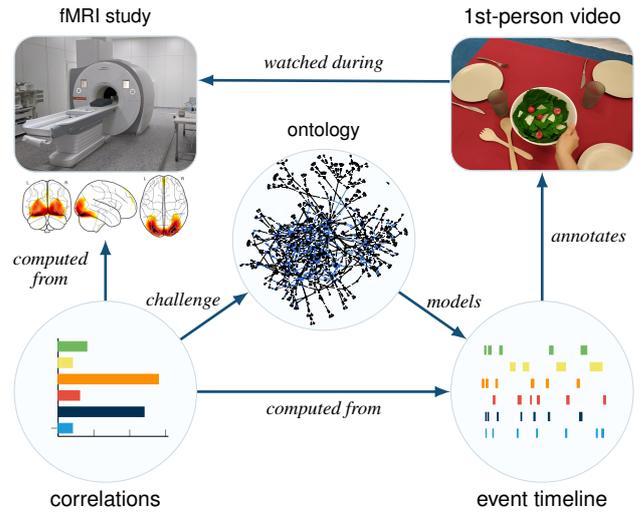}
    \caption{Correlations between neuronal components and event segmentation are used to challenge definitions in an existing ontology.}
    \label{fig:abstract}
\end{figure}

If interoperability between very different systems and/or knowledge transfer between different expert communities is needed, then ontologies are a useful tool towards that goal. An ontology is a collection of axioms in a formal, machine-readable, language which defines terms and makes explicit a conceptualization shared within a community of people.
In the case of NEEM narratives, it is the \emph{Socio-physical Model of Activities }(SOMA)~ontology~\cite{bessler21soma}
which defines the semantic categories which constitute everyday activities and the relationships between entities participating in such activities. As such, \SOMA~models knowledge about activities both from a robotics perspective, but also aims to model such knowledge as it would be used by humans to organize their own behavior.

It is thereby crucial that \SOMA, 
as the basis for the classification of both human and robot behavior and perception, correctly represents and interfaces the experience of both.
For the case of robotic agents, this has already been demonstrated~\cite{koralewski19specialization}.
On the human side, a correct representation is arguably ensured by a learned cognitive model that represents the rules for segmentation of everyday behavior into distinctive categories. There is a correlation with the underlying experience stored in dynamically organized networks of brain areas, due to the human ability for self-reflection.
However, the process of rule building, itself based on complex brain activity, might introduce different layers of abstraction that obscure aspects of the neuronal activity underlying our experience of everyday activities. 
It would be beneficial to access the neuronal level to ascertain whether \SOMA~reflects the underlying processing of these events. 

Given the prerequisite as correct that \SOMA~is considered valid for classifying human experience on a behavioral as well as neuronal level, further analysis directed towards neuronal preferences for certain event classes or levels and a subsequent integration of these neuronal characteristics into the \SOMA~ontology would be a direct proof of the robot cognitive architecture’s being inspired by basic principles of human information processing.
We thus decided to investigate the \SOMA~ontology on a neuronal level
via \emph{functional Magnetic Resonance Imaging} (fMRI) to capture brain activity of participants who experienced everyday activities as covered by the \SOMA~ontology.
Since our participants were required to remain in a stationary position inside the MR-scanner, data collection made use of the human brain's ability to simulate environmental interactions.

This procedure refers to the basic assumption that brain activity patterns in the absence of external stimuli through motor imagery~\cite{kosslyn_seeing_1987, jeannerod_representing_1994}, follow similar psycho-physical rules as realized during action execution~\cite{bakker_motor_2007, frak_orientation_2001} and lead to comparable patterns of brain activation in perceptual~\cite{ pearson_human_2019}, as well as motor-related brain networks~\cite{gerardin_partially_2000, roth_possible_1996, stippich_somatotopic_2002}.
If actions are not only imagined but observed, the human equivalent of the mirror neuron system is hypothesized to be recruited as an additional system for mapping observed actions into the subject's own motor representation~\cite{rizzolatti_neurophysiological_2001}. Both motor imagery and action observation are thought to arise from a complex neuronal network for learning, maintenance, and refinement of motor actions through a synergy of execution and simulation~\cite{hesslow_conscious_2002, jeannerod_neural_2001}. This network allows for a close approximation of neuronal correlates of real-world interaction when measuring fMRI participants, especially through use of immersive stimuli with bio-mechanical action representations that have a high degree of ecological validity and correspondence to the observing participants and their motor capabilities~\cite{stevens_new_2000, tai_human_2004}. 

Despite lack of physical engagement in the presented actions during fMRI scanning, it was therefore hypothesized that, given an appropriate level of immersion, brain activity patterns would resemble those present in the real-world experience and therefore allow for a neuronal validation of \SOMA~ontology through correlation with the spatio-temporal characteristics of its ontological classes as depicted in figure~\ref{fig:abstract}. This would contribute to not only \SOMA~development but ontology development in general by evaluation of an ontology's neuronal validity and potential detection of preferential processing of event categories, enabling new insights for ontology design.

\section{Related Work}
The work presented here relates to the field of ontology and knowledge engineering as well as to the field of neuroimaging studies during naturalistic viewing.
In the following, we will relate our work to the corpus of existing literature in these two fields.

\subsection{Knowledge Engineering and Ontology}

Information retrieval/organization have motivated ontology creation for medicine and bioinformatics.
Examples are the \emph{Neuroscience Information Framework}~\cite{Imam2012} and the ontologies used for data integration in the \emph{Human Brain Project}~\cite{bjerke2018}.
A survey of ontologies for the study of Alzheimer's is provided in~\cite{gomez2019ontologies}. Mercier et al. present steps towards an ontological model, informed by cognitive research, of how a human learns to solve a computational problem~\cite{mercier2021formalizing}; their model is also able to predict learner behavior to some degree. 

Ontologies have also been applied to fields such as robotics.
As an example, the SOMA ontology~\cite{bessler21soma} defines concepts for autonomous robots to use while performing everyday activities in the home.
A broad survey on robotics ontologies is provided by~\cite{Olivares19}.

Some works~\cite{McCaffrey2016, anderson2014after, beam2021data} make a loose distinction between top-down, knowledge-driven and bottom up, data-driven, methods to uncover scientifically-relevant entities. 
Bottom-up approaches are less affected by expert preconceptions and can recover robust patterns~\cite{beam2021data}, but may confuse dimensionality reduction artifacts with causal mechanisms~\cite{McCaffrey2016}. Ontologies are developed by a mix of empirical and conceptual issues related to what kinds of entities and questions may be relevant. A summary of this debate can be found in~\cite{anderson2014after} and~\cite{McCaffrey2016}.

Our work here is itself a hybrid top-down/bottom-up approach, in that we start from an ontology developed top-down for robotic actions but compare with human neurological data to ascertain what distinctions between actions humans find relevant. 

\subsection{Brain Patterns of Naturalistic Viewing}
Functional brain imaging has traditionally favoured simple, static stimuli, but the use of complex, dynamic ones may be crucial for analysing the brain in its most natural state (see:~\cite{eickhoff_towards_2020, leopold_studying_2020, simony_analysis_2020}). Since the \SOMA~ontology is meant to describe the dynamic processes carried out by human- and robot agents, neuronal correlates would have to stem from such kinds of stimuli in order to assess its neuronal validity.

Presentation of complex dynamic stimuli during fMRI measurements was shown to be feasible via video with a high degree of inter-participant spatio-temporal correlation of brain activity~\cite{hasson_intersubject_2004, byrge_video-evoked_2021}, leading to insights into the neuronal correlates of event and event boundary perception (e.g.,~\cite{zacks_human_2001}). It was further shown that machine learning models could accurately predict the perception of semantic categories from data recorded during video presentation~\cite{huth_decoding_2016}.

For initial dimensionality reduction of fMRI recordings, data driven models such as \emph{Independent Component Analysis} (ICA) are used for clustering brain activity into distinct networks through blind source separation~\cite{mckeown2003independent}. For fMRI data recorded under natural viewing conditions, it was used to subdivide whole-brain activity into spatio-temporal components with characteristic activity time-courses. The association of network activity underlying these components with presented stimuli was analyzed through inter-subject correlation of time-courses, resulting in differentiation of relevant components from non-stimulus related activity and artifacts~\cite{bartels_chronoarchitecture_2004}.

A recording and analysis of neuronal dynamics of everyday activities, which supported the possibility of detecting event dependent allocation of brain activity via \emph{General Linear Model} (GLM) and ICA in the scope of the present research framework was carried out by Ahrens2021~\cite{ahrens_neuronal_2021}. 
The study's ICA analysis thereby directly correlated component time-courses with semantically annotated events. Results indicated a set of components common to participants whose activity exhibited such correlations with an additional inter-component preference depending on the broad classification of the event. These results were however too vague to be used to study human understanding of events.

An updated ICA analysis based on the same data set was thus carried out for the scope of this paper that followed a more stringent ruleset, including multiple comparison correction and a strict focus on stable correlations that were shared amongst participants in order to achieve a more focused set of results.

\section{fMRI study}
\subsection{Experimental Design \& Annotation}
\begin{figure}[ht]
    \centering
    \input{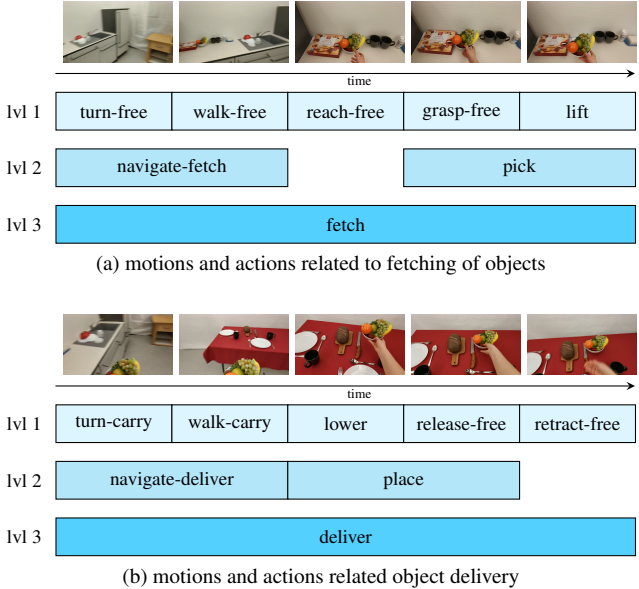}
    \caption{Exemplary visualization of events on annotation levels 1-3.}
    \label{fig:annot_example}
\end{figure}

Stimulus material consisted of ten 1st-person videos of 29 – 105 s duration, recorded via a head mounted camera. As demonstrated in the top rows of figure~\ref{fig:anot_overview_fetch}~\&~\ref{fig:anot_overview_deliver}, the videos depicted table setting activities in the \emph{Cognitive Systems Lab} (CSL) \& the \emph{Institute for Artificial Intelligence} (IAI)) of the EASE CRC at the University of Bremen, integrating the main venues for robot- (IAI) and human- (CSL) activity research into the analysis and representing the project's larger focus on table-setting activities. 

The videos covered the fetching of objects from a source area and their subsequent deliverance to a target area for placement in table-setting scenarios. Each scenario was split into a video pair with one part showing the setting of dishes \& cutlery and the other part dealing with setting of food \& drink items, resulting in four video classes (IAI-dishes, IAI-food, CSL-dishes, CSL-food). Three scenarios were recorded in the CSL and two in the IAI.
Both venues differed in factors such as environmental complexity, with the IAI featuring a realistic kitchen environment versus the sparser environment of the CSL, consisting of two tables acting as source and target areas. With the additional split into dish- and food-videos, this allowed for the analysis of contextual effects on measured brain activity and resulting changes in correlation to ontology classes. 

Interactions were carried out in a realistic manner, including single- and two-handed movements, while ensuring that all actions remained traceable and recognizable for the viewer.
Videos were embedded into an experimental design consisting of two sequences (A \& B) of presented table-setting videos and interspersed resting periods. Participants were instructed to watch the table-setting presentations attentively and imagine being the acting protagonist.
Both sequences covered ten 1st-person videos, albeit in a switched order with respect to presentation of food \& drink and dishes \& cutlery parts. Stimuli were presented in a counter-balanced order: half of the study’s participants were first presented with sequence A then sequence B, the other half in a reversed order. At the end of a measurement, every participant thus was presented each video twice.

Videos were annotated with EASELAN~\cite{meier_comparative_2019}, a modification of the ELAN~\cite{ELAN} software by the Max Planck Institute for Psycholinguistics.
The process was carried out according to a predefined subset of \SOMA~events  for human activity description developed in collaboration between the human activity and ontology subgroups of EASE CRC. The subset consisted of event categories, nested into annotation levels of increasing length and complexity. Examples of this are depicted in figure~\ref{fig:annot_example}. 

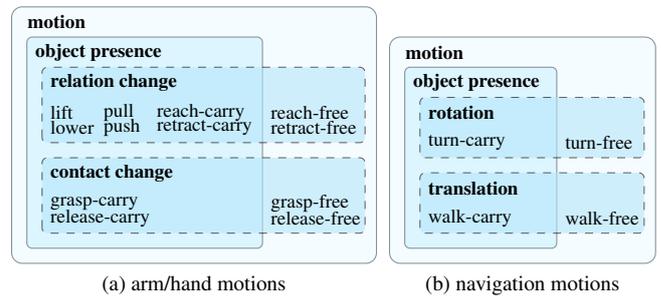
\begin{figure}
\centering
    \subfloat[arm/hand motions]{
    \centering
        \begin{tikzpicture}
            \draw[draw=motionBack!60!black,fill=motionBack!30!white,rounded corners=4pt] (0,0) rectangle (4.8,3.4);
            \draw[draw=motionBack!60!black,fill=motionBack,rounded corners=2pt] (0.2,0.2) rectangle (3.3,3);
            \draw[draw=black,fill=actionBack,opacity=0.6,rounded corners=2pt,dashed] (0.4,1.6) rectangle (4.65,2.6);
            \draw[draw=black,fill=actionBack,opacity=0.6,rounded corners=2pt,dashed] (0.4,0.4) rectangle (4.65,1.4);

            \node[anchor=west] at (0.1,3.2) {\scriptsize\textbf{motion}};
            \node[anchor=west] at (0.2,2.8) {\scriptsize\textbf{object presence}};
            \node[anchor=west] at (0.4,2.4) {\scriptsize\textbf{relation change}};
            \node[anchor=west] at (0.4,1.2) {\scriptsize\textbf{contact change}};
                
            \node[anchor=west] at (3.3,2) {\scriptsize{reach-free}};
            \node[anchor=west] at (3.3,1.8) {\scriptsize{retract-free}};
                
            \node[anchor=west] at (1.8,2) {\scriptsize{reach-carry}};
            \node[anchor=west] at (1.8,1.8) {\scriptsize{retract-carry}};
            \node[anchor=west] at (0.4,2) {\scriptsize{lift}};
            \node[anchor=west] at (0.4,1.8) {\scriptsize{lower}};
            \node[anchor=west] at (1.1,2) {\scriptsize{pull}};
            \node[anchor=west] at (1.1,1.8) {\scriptsize{push}};
                
            \node[anchor=west] at (3.3,0.8) {\scriptsize{grasp-free}};
            \node[anchor=west] at (3.3,0.6) {\scriptsize{release-free}};
                
            \node[anchor=west] at (0.4,0.8) {\scriptsize{grasp-carry}};
            \node[anchor=west] at (0.4,0.6) {\scriptsize{release-carry}};

        \end{tikzpicture}
        \label{fig:anot_motion_arms}
    }%
    \subfloat[navigation motions]{
    \centering
        \begin{tikzpicture}
            \draw[draw=motionBack!60!black,fill=motionBack!30!white,rounded corners=4pt] (0,0) rectangle (3.5,3);
            \draw[draw=motionBack!60!black,fill=motionBack,rounded corners=2pt] (0.2,0.2) rectangle (2.2,2.6);
            \draw[fill=actionBack,opacity=0.6,rounded corners=2pt,dashed] (0.4,1.4) rectangle (3.35,2.2);
            \draw[fill=actionBack,opacity=0.6,rounded corners=2pt,dashed] (0.4,0.4) rectangle (3.35,1.2);
    
            \node[anchor=west] at (0.1,2.8) {\scriptsize\textbf{motion}};
            \node[anchor=west] at (0.2,2.4) {\scriptsize\textbf{object presence}};
            \node[anchor=west] at (0.4,2) {\scriptsize\textbf{rotation}};
            \node[anchor=west] at (0.4,1) {\scriptsize\textbf{translation}};
    
            \node[anchor=west] at (2.2,1.6) {\scriptsize{turn-free}};
            \node[anchor=west] at (2.2,0.6) {\scriptsize{walk-free}};
            
            \node[anchor=west] at (0.4,1.6) {\scriptsize{turn-carry}};
            \node[anchor=west] at (0.4,0.6) {\scriptsize{walk-carry}};
        \end{tikzpicture}
        \label{fig:anot_motion_nav}
        }%
    
    \caption{Atomic motion-level annotation scheme.}
    \label{fig:anot_motion_all}
\end{figure}

The lowest annotation level (level 1) consisted of the simplest events in 1st-person videos of everyday activities that are still distinguishable using (combinations of) \SOMA~motion concepts. These atomic motions are listed in figure~\ref{fig:anot_motion_all}. They were grouped as either arm/hand motions related to object interaction or motions related to the body’s navigation in space. A further distinction was made based on whether an object was held while a motion was performed.

This was marked by 'carry' and 'free'.
For the motions of grasp and release, it describes whether additional objects were held while performing the respective motion, i.e., grasp-carry indicated that one or more objects were already held while an additional one was grabbed, release-carry indicated that after a release of one object, one or more remained in hand. Other distinctions between motions made by \SOMA~relate to details of a motion. For example, while 'reach' and 'push' involved similar movements of the hand relative to the body, they involved different force profiles, different contacts, and different states of control over a manipulated object.

\begin{figure}
\centering
\centering
    \subfloat[arm/hand actions]{
        \begin{tikzpicture}
            \draw (0,5) node[fill=motionBack,anchor=west,minimum height=0.5cm,minimum width=1.625cm,draw] (grasp_pick) {\scriptsize{grasp}};
            \draw (1.625,5) node[fill=motionBack,anchor=west,minimum height=0.5cm,minimum width=1.625cm,draw] (lift) {\scriptsize{lift}};
            \draw (4,5) node[fill=motionBack,anchor=west,minimum height=0.5cm,minimum width=1.625cm,draw] (lower) {\scriptsize{lower}};
            \draw (5.625,5) node[fill=motionBack,anchor=west,minimum height=0.5cm,minimum width=1.625cm,draw] (release_place) {\scriptsize{release}};
    
            \draw (0,3.75) node[fill=motionBack,anchor=west,minimum height=0.5cm,minimum width=1.625cm,draw] (grasp_open) {\scriptsize{grasp}};
            \draw (1.625,3.75) node[fill=motionBack,anchor=west,minimum height=0.5cm,minimum width=1.625cm,draw] (pull) {\scriptsize{pull}};
            \draw (4,3.75) node[fill=motionBack,anchor=west,minimum height=0.5cm,minimum width=1.625cm,draw] (push) {\scriptsize{push}};
            \draw (5.625,3.75) node[fill=motionBack,anchor=west,minimum height=0.5cm,minimum width=1.625cm,draw] (release_close) {\scriptsize{release}};
            
            \draw (0,2.5) node[fill=motionBack,anchor=west,minimum height=0.5cm,minimum width=1.625cm,draw] (grasp_switch) {\scriptsize{grasp}};
            \draw (1.625,2.5) node[fill=motionBack,anchor=west,minimum height=0.5cm,minimum width=1.625cm,draw] (release_switch) {\scriptsize{release}};
            
            \draw (0,4.5) node[fill=actionBack,anchor=west,minimum height=0.5cm,minimum width=3.25cm,draw] (pick) {\scriptsize{pick}};
            \draw (4,4.5) node[fill=actionBack,anchor=west,minimum height=0.5cm,minimum width=3.25cm,draw] (place) {\scriptsize{place}};
    
            \draw (0,3.25) node[fill=actionBack,anchor=west,minimum height=0.5cm,minimum width=3.25cm,draw] (open) {\scriptsize{open}};
            \draw (4,3.25) node[fill=actionBack,anchor=west,minimum height=0.5cm,minimum width=3.25cm,draw] (close) {\scriptsize{close}};
            
            \draw (0,2) node[fill=actionBack,anchor=west,minimum height=0.5cm,minimum width=3.25cm,draw] (switch) {\scriptsize{switch-hands}};
        
        \end{tikzpicture}
        \label{fig:anot_action_arms}
    }\\
    \subfloat[navigation actions]{
        \begin{tikzpicture}
            \draw (0,5) node[fill=motionBack,anchor=west,minimum height=0.5cm,minimum width=1.625cm,draw] (turn_free) {\scriptsize{turn-free}};
            \draw (1.625,5) node[fill=motionBack,anchor=west,minimum height=0.5cm,minimum width=1.625cm,draw] (walk_free) {\scriptsize{walk-free}};
            \draw (4,5) node[fill=motionBack,anchor=west,minimum height=0.5cm,minimum width=1.625cm,draw] (turn_carry) {\scriptsize{turn-carry}};
            \draw (5.625,5) node[fill=motionBack,anchor=west,minimum height=0.5cm,minimum width=1.625cm,draw] (walk_carry) {\scriptsize{walk-carry}};
    
            \draw (0,4.5) node[fill=actionBack,anchor=west,minimum height=0.5cm,minimum width=3.25cm,draw] (nav_fetch) {\scriptsize{navigate to fetch}};
            \draw (4,4.5) node[fill=actionBack,anchor=west,minimum height=0.5cm,minimum width=3.25cm,draw] (nav_deliver) {\scriptsize{navigate to deliver}};
        \end{tikzpicture}
        \label{fig:anot_action_nav}
        }
    \caption{Action-level annotation scheme, including constituting motions.}
    \label{fig:anot_action_all}
\end{figure}
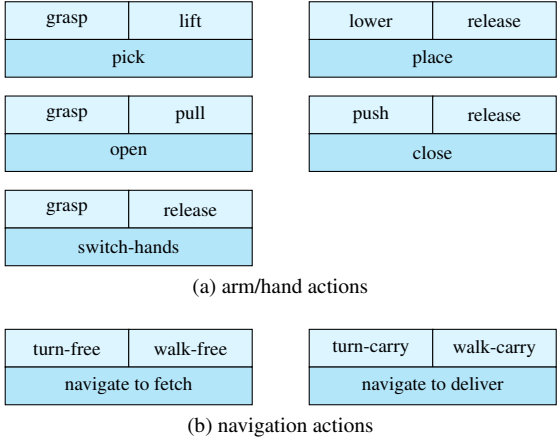
The level above atomic motions was the action level (level 2). Actions are performed to reach certain goals and are comprised of a set of motions. For the video annotation we used the actions shown in figure~\ref{fig:anot_action_all}. Five action categories were derived from arm/hand motions that covered picking and placing of objects as well as opening and closing of doors and drawers and switching an object between hands. For navigation, we covered two actions. 'Navigate to fetch' represented the navigation motions, usually a rotation followed by a translation, needed to reach the source area, whereas 'navigate to deliver' covered the respective motions for arriving at the target area.

The level above the action level was the situation level (level 3). For our annotation it consisted of two large-grained time-frames of 'fetch' and 'deliver'. 'Fetch' covered all navigation and object interaction motions that lead to the fetching of item from a source area. During 'deliver', objects were transported to the target area and set down. Examples are shown in the bottom rows of figure~\ref{fig:anot_overview_fetch} \& ~\ref{fig:anot_overview_deliver}. 

Additionally, we included an exploratory group level (level 4) by grouping the atomic motions based on their affiliation to overlapping classes and sub-classes.
In most general terms, we grouped motions based on their affiliation to either arm/hand based object interaction motions listed in figure~\ref{fig:anot_motion_arms} and navigation motions, listed in figure~\ref{fig:anot_motion_nav}. Within those groups, we subdivided based on 'object presence', describing whether an object was present during motion. 
For object interaction, we further introduced the categories of 'relation change', describing motions that resulted in change of the spatial relation between hand and body as well as 'contact change', describing motions that either established or broke a contact between hand and object. 
For navigation motions, we further introduced the categories 'rotation', describing motions in which the body/view is rotated either clock- or anticlockwise and 'translation', describing a forward movement of the body/view in space.
Groups were built out of all class affiliations except for navigation motions. Here, no grouping based on object presence was performed since this category was covered by the actions (level 2) of ’navigate to fetch’ and ’navigate to deliver’.

Overall, this resulted in a maximum of 16 motion-, 7 action-, 2 situation- and 8 group-categories per video, leading to an overall maximum number of 33 event categories.

\subsection{Participants \& Data Acquisition}
Thirty participants (21 identifying as female) with a mean age of 23.3 years (SD = 4.54 years) were recruited from the campus of the University of Bremen.
All subjects were right-handed, healthy by own accord, and naïve to the experiment before arriving at the laboratory. Prior to the scan session, participants were given a brief tutorial, including presentation of two short videos similar to those shown during the experiment. All stimuli were shown through a mirror system attached to the head coil. Videos were displayed in a rectangle over a black background.
Imaging data were acquired with a Siemens 3 Tesla MAGNETOM Skyra full body scanner. FMRI data were recorded via T2*-weighted multi-band EPI with acceleration factor = 3, TR = 1.1 s, TE = 30 ms, matrix size = 64x64x45 voxels and voxel size = 3x3x3 mm. After completion of the functional recording, a T1-weighted structural scan with matrix size = 255x265x265 voxels and voxel size = 1x1x1 mm was performed.

\subsection{Data Processing \& Analysis}
FMRI data were taken from original recordings made for Ahrens2021~\cite{ahrens_neuronal_2021} which were preprocessed in the \emph{Statistical Parametric Mapping toolbox V12}~\cite{SPM12} in \emph{MATLAB V2018b}~\cite{MATLAB}, via subsequent slice-time correction, realignment, coregistration and normalization to standard brain template. Data sets were spatially smoothed with an 8 mm FWHM Gaussian kernel.
For each participant, data were separated into blocks temporally corresponding to the presented video- and resting trials via a script taking the \emph{Hemodynamic Response Function} (HRF) into account. For partition of participants' brain activity into spatially independent sources, a spatial ICA was performed on a group level with the \emph{Group ICA of fMRI MATLAB Toolbox V4.0c}~\cite{GIFT} over all participants’ blocks through an infomax algorithm with an automatic estimation of components numbers. An ICASSO analysis was employed to ascertain the stability of calculated components over repeated calculations~\cite{himberg_validating_2004}. This resulted in classification of brain activity into 15 components common to all participants over all video- and resting trials.

\SOMA~association of brain activity in component maps was calculated trough Spearman rank correlation between the subject-specific component activity time-course during each video trial and the respective HRF-convolved stimulus timings of the presented sequences of events of a respective category in levels 1-4.
Deviating from Ahrens2021~\cite{ahrens_neuronal_2021}, correlation coefficients had to pass a statistical threshold of $p \leq 0.05$, with a Holm-Bonferroni correction~\cite{holm_simple_1979} for multiple comparisons. The remaining significant correlations further had to proof stable over subsequent presentations of a respective video for each participant. Finally, each stable correlation needed to be found in two or more participants to contribute to the resulting data-set. Components that exhibited such stable and shared correlations were listed and analysed based on their spatial and temporal characteristics.

\section{Results}
\subsection{Grading of \SOMA~Association}
\begin{figure}
    \centering
\centering
    \subfloat[Number of participants with significant correlation of brain activity and annotated events per component.]{
        \pgfplotstableread[col sep=comma,]{./data/taskRelDoublePosCSV.txt}\taskRelTable
        \begin{tikzpicture}
            \begin{axis}[
                        ybar,
                        scale only axis,
                        width=0.8\columnwidth,
                        height=2.25cm,
                        ymin = 0,
                        xtick=data,
                        xticklabels from table={\taskRelTable}{intercomCompSortedNoZero},
                        xlabel={component no.},
                        ylabel={no. of participants }]
                        \addplot [black,fill=barMainCol] table[y=intercomNumSortedNoZero]{\taskRelTable};
                    \end{axis}
        \end{tikzpicture}
        \label{fig:soma_relevant_subjNo}
    }\\
    \subfloat[Number of event categories with significant correlations. Exclusive categories exhibited significant correlation for only a single participant. Shared categories were found in at least two participants A participant could show significant correlation in more than one category for each combination of video and component]{
        \pgfplotstableread[col sep=comma,]{./data/taskRel_sumCategs.txt}\taskRelSumCategsTable
            \begin{tikzpicture}
                \begin{axis}[
                            ybar stacked,
                            scale only axis,
                            legend pos=north west,
                            legend columns=1,
                            xtick=data,
                            width=0.8\linewidth,
                            height=2.25cm,
                            ymin=0,
                            area legend,
                            xlabel={component no.},
                            ylabel={no. of event categories},
                            xticklabels from table={\taskRelSumCategsTable}{tempSumCategs_comSorted}]
                       
                    \addplot[black,fill=barMainCol] table[x=xPos, y=tempSumCategs_valSorted_minNum]{\taskRelSumCategsTable};
                    \addplot[black,fill=none] table[x=xPos, y=tempSumCategs_valSorted_nonEmptyDiffMin ]{\taskRelSumCategsTable};
                    \legend{shared, exclusive}
                \end{axis}  
        \end{tikzpicture}
        \label{fig:soma_relevant_catNo}
        }
    \caption[\SOMA~relatedness]{\SOMA~association grading of resulting ICA components.}
    \label{fig:soma_interComp}
\end{figure}
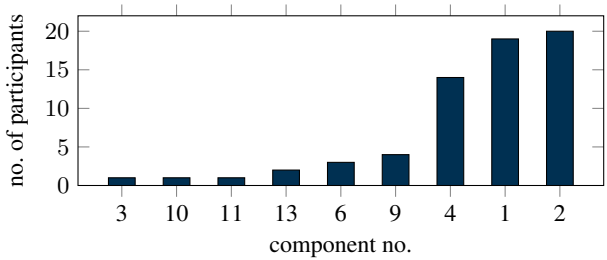
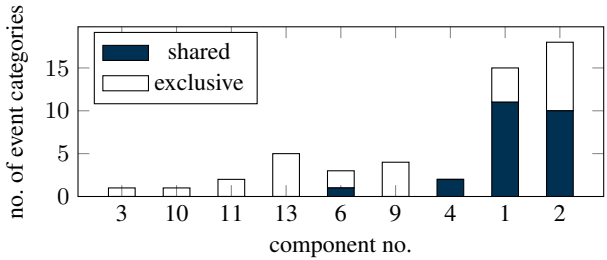

In nine of fifteen components, at least one out of thirty participants exhibited significant and stable correlations between brain activity within the represented network and at least one annotated event category, as shown in figure~\ref{fig:soma_relevant_subjNo}. While for components 3, 10, and 11, only one participant was found, two were found in component 13, three in component 6 and four in component 9. The highest numbers of participants were found in component 4 with 14 participants, component 1 with 19 participants and component 2 with 20 participants.
The number of event categories with significant and stable correlations for each component is depicted in figure~\ref{fig:soma_relevant_catNo}, split into categories exclusive to a particular participant and categories shared between at least two participants. Out of the nine components, only four exhibited shared event categories. In component 6, one such category was found, while two were present in component 4. The highest number of categories were present in component 2 with ten categories and component 1 with eleven categories.

\subsection{ Spatial Brain Activation Maps of Components with shared \& stable \SOMA~Association}
Brain activity patterns associated with component 1 (fig.~\ref{fig:ica_maj_comp01_spat}) were primarily found in the temporal- and occipital lobes, with largest activation in the lateral occipital cortex and extending into the medioventral occipital cortex. In the temporal lobe, the component encompassed the fusiform gyrus. Smaller clusters were also found in the inferior parietal lobule as well as the cerebellum.

Activity associated with component 2 (fig.~\ref{fig:ica_maj_comp02_spat}) occurred in the parietal lobe with clusters the superior- and inferior parietal lobule, the precuneus and postcentral gyrus. It extended into parts of the lateral occipital cortex, the middle temporal gyrus and inferior temporal gyrus. In the frontal lobe, small clusters were situated in the superior- and middle frontal gyrus, precentral gyrus, and paracentral lobule.

The largest activity cluster for component 4 (fig.~\ref{fig:ica_maj_comp04_spat}) was found spanning areas of the medioventral occipital cortex with smaller clusters in the lateral occipital cortex. It expanded into the superior parietal lobule and the precuneus in the parietal lobe, and the fusiform gyrus and parahippocampal gyrus in the temporal lobe. In the limbic lobe, a small cluster was found in cingulate gyrus.

Activity in component 6 (fig.~\ref{fig:ica_maj_comp06_spat}) was located in the postcentral gyrus of the parietal lobes as well as the superior temporal gyri of the temporal lobe. In the frontal lobe, the component comprised the paracentral lobule and precentral- as well as superior frontal gyri.

\begin{figure}
    \centering
    \input{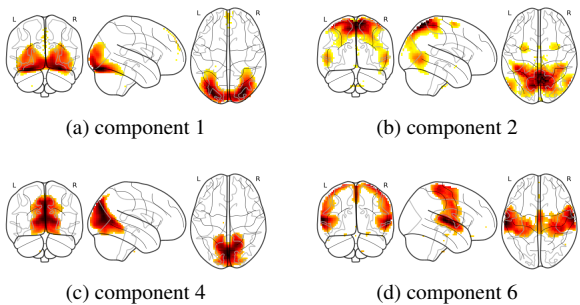}
    \caption{Spatial maps of resulting ICA components that exhibited shared and stable correlations with events of the ontology.}
    \label{fig:ica_main_spat}
\end{figure}

\subsection{Intra-Component Correlations}
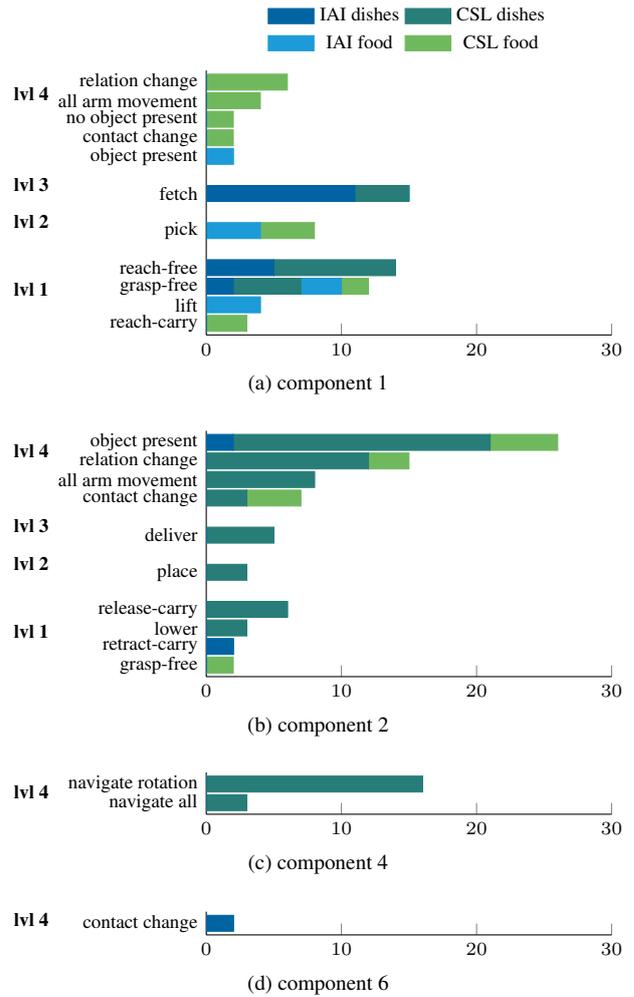
\begin{figure}[t]
    \subfloat[component 1]{
        \begin{tikzpicture}
        \pgfplotstableread[col sep=comma,]{./data/intra/comp1_allHits_gap.txt}\allHitsCurrCompTable
        \begin{axis}[
            xbar stacked,
            legend style={
            legend columns=2,
                at={(0.5,1.03)},anchor=south,
                draw=none
            },
            ytick=data,
            axis y line*=none,
            axis x line*=bottom,
            tick label style={font=\scriptsize},
            legend style={font=\scriptsize},
            label style={font=\scriptsize},
            xtick={0,10,20,30},
            width=.8\columnwidth,
            bar width=\horzBarWidth,
            xmin=0,
            xmax=\horzBarXlim,
            area legend,
            y=\horzBarY,
            enlarge y limits={abs=0.625},
            yticklabels from table={\allHitsCurrCompTable}{categs_sorted}
        ]
        \addplot[barIAIdishes,fill=barIAIdishes] table[x=sumHits_iaiDishes_sorted, y=yPos]{\allHitsCurrCompTable};
        \addplot[barCSLdishes,fill=barCSLdishes] table[x=sumHits_cslDishes_sorted, y=yPos]{\allHitsCurrCompTable};
        \addplot[barIAIfood,fill=barIAIfood] table[x=sumHits_iaiFood_sorted, y=yPos]{\allHitsCurrCompTable};
        \addplot[barCSLfood,fill=barCSLfood] table[x=sumHits_cslFood_sorted, y=yPos]{\allHitsCurrCompTable};
        \legend{IAI dishes,CSL dishes,IAI food,CSL food}
        \end{axis}
        \draw (-2.3,3.2) node[font={\scriptsize}] (lvl4) {\textbf{lvl 4}};
        \draw (-2.3,2) node[font={\scriptsize}] (lvl3) {\textbf{lvl 3}};
        \draw (-2.3,1.5) node[font={\scriptsize}] (lvl2) {\textbf{lvl 2}};
        \draw (-2.3,0.6) node[font={\scriptsize}] (lvl1) {\textbf{lvl 1}};
        \end{tikzpicture}
        \label{fig:events_comp01}
    }\\
    \subfloat[component 2]{
        \begin{tikzpicture}
        \pgfplotstableread[col sep=comma,]{./data/intra/comp2_allHits_gap.txt}\allHitsCurrCompTable
        \begin{axis}[
            xbar stacked,
            legend style={
            legend columns=2,
                at={(xticklabel cs:0.5)},
                anchor=north,
                draw=none
            },
            ytick=data,
            axis y line*=none,
            axis x line*=bottom,
            tick label style={font=\scriptsize},
            legend style={font=\scriptsize},
            label style={font=\scriptsize},
            xtick={0,10,20,30},
            width=.8\columnwidth,
            bar width=\horzBarWidth,
            xmin=0,
            xmax=\horzBarXlim,
            area legend,
            y=\horzBarY,
            enlarge y limits={abs=0.625},
            yticklabels from table={\allHitsCurrCompTable}{categs_sorted}
        ]
        \addplot[barIAIdishes,fill=barIAIdishes] table[x=sumHits_iaiDishes_sorted, y=yPos]{\allHitsCurrCompTable};
        \addplot[barCSLdishes,fill=barCSLdishes] table[x=sumHits_cslDishes_sorted, y=yPos]{\allHitsCurrCompTable};
        \addplot[barIAIfood,fill=barIAIfood] table[x=sumHits_iaiFood_sorted, y=yPos]{\allHitsCurrCompTable};
        \addplot[barCSLfood,fill=barCSLfood] table[x=sumHits_cslFood_sorted, y=yPos]{\allHitsCurrCompTable};
        \end{axis}  
        \draw (-2.3,3) node[font={\scriptsize}] (lvl4) {\textbf{lvl 4}};
        \draw (-2.3,2) node[font={\scriptsize}] (lvl3) {\textbf{lvl 3}};
        \draw (-2.3,1.5) node[font={\scriptsize}] (lvl2) {\textbf{lvl 2}};
        \draw (-2.3,0.6) node[font={\scriptsize}] (lvl1) {\textbf{lvl 1}};
        \end{tikzpicture}
        \label{fig:events_comp02}
    }\\
    \subfloat[component 4]{
        \begin{tikzpicture}
        \pgfplotstableread[col sep=comma,]{./data/intra/comp4_allHits_gap.txt}\allHitsCurrCompTable
        \begin{axis}[
            xbar stacked,
            legend style={
            legend columns=2,
                at={(xticklabel cs:0.5)},
                anchor=north,
                draw=none
            },
            ytick=data,
            axis y line*=none,
            axis x line*=bottom,
            tick label style={font=\scriptsize},
            legend style={font=\scriptsize},
            label style={font=\scriptsize},
            xtick={0,10,20,30},
            width=.8\columnwidth,
            bar width=\horzBarWidth,
            xmin=0,
            xmax=\horzBarXlim,
            area legend,
            y=\horzBarY,
            enlarge y limits={abs=0.625},
            yticklabels from table={\allHitsCurrCompTable}{categs_sorted}
        ]
        \addplot[barIAIdishes,fill=barIAIdishes] table[x=sumHits_iaiDishes_sorted, y=yPos]{\allHitsCurrCompTable};
        \addplot[barCSLdishes,fill=barCSLdishes] table[x=sumHits_cslDishes_sorted, y=yPos]{\allHitsCurrCompTable};
        \addplot[barIAIfood,fill=barIAIfood] table[x=sumHits_iaiFood_sorted, y=yPos]{\allHitsCurrCompTable};
        \addplot[barCSLfood,fill=barCSLfood] table[x=sumHits_cslFood_sorted, y=yPos]{\allHitsCurrCompTable};
        \end{axis}  
        \draw (-2.3,0.3) node[font={\scriptsize}] (lvl4) {\textbf{lvl 4}};
        \end{tikzpicture}
        \label{fig:events_comp04}
    }\\
    \subfloat[component 6]{
        \begin{tikzpicture}
        \pgfplotstableread[col sep=comma,]{./data/intra/comp6_allHits_gap.txt}\allHitsCurrCompTable
        \begin{axis}[
            xbar stacked,
            legend style={
            legend columns=2,
                at={(xticklabel cs:0.5)},
                anchor=north,
                draw=none
            },
            ytick=data,
            axis y line*=none,
            axis x line*=bottom,
            tick label style={font=\scriptsize},
            legend style={font=\scriptsize},
            label style={font=\scriptsize},
            xtick={0,10,20,30},
            width=.8\columnwidth,
            bar width=\horzBarWidth,
            xmin=0,
            xmax=\horzBarXlim,
            area legend,
            y=\horzBarY,
            enlarge y limits={abs=0.625},
            yticklabels from table={\allHitsCurrCompTable}{categs_sorted}
        ]
        \addplot[barIAIdishes,fill=barIAIdishes] table[x=sumHits_iaiDishes_sorted, y=yPos]{\allHitsCurrCompTable};
        \addplot[barCSLdishes,fill=barCSLdishes] table[x=sumHits_cslDishes_sorted, y=yPos]{\allHitsCurrCompTable};
        \addplot[barIAIfood,fill=barIAIfood] table[x=sumHits_iaiFood_sorted, y=yPos]{\allHitsCurrCompTable};
        \addplot[barCSLfood,fill=barCSLfood] table[x=sumHits_cslFood_sorted, y=yPos]{\allHitsCurrCompTable};
        \end{axis}
        \draw (-2.3,0.2) node[font={\scriptsize}] (lvl4) {\textbf{lvl 4}};
        \end{tikzpicture}
        \label{fig:events_comp06}
    }
    \caption{Number of shared and stable correlations for neuronal networks represented by resulting ICA components.}
    \label{fig:events_compRelevant}
\end{figure}

The number of shared significant stable correlations for all \SOMA~associated components is depicted in figure~\ref{fig:events_compRelevant} via bar graphs for all four video classes.
Figure~\ref{fig:events_comp01} depicts shared stable correlations for component 1 ranging over all annotation levels.
On level 1, the ICA component correlated with four motion categories (33 correlations in total). Its highest number is for 'reach-free' (14 correlations, 9 CSL-dishes, 5 IAI-dishes), followed by 'grasp-free' (12 correlations, 5 CSL-dishes, 3 IAI-food, 2 IAI-dishes, 2 CSL-food), 'lift' (4 correlations in IAI-food) and 'reach-carry' (3 correlations in CSL-food). 
On level 2, it correlated with the action category of 'pick' (8 correlations, 4 CSL-food, 4 IAI-food). 
On level 3, it correlated with the situation category of 'fetch' (15 correlations, 11 IAI-dishes, 4 CSL-dishes). 
On level 4, it correlated with five exploratory groups (16 correlations in total). Highest correlation is with 'relation change' (6 correlations in CSL-food), followed by 'all arm movement' (4 correlations in CSL-food) and the categories of 'no object present' (2 correlations in CSL-food), 'contact change' (2 correlations in CSL-food) and 'object present' (2 correlations in IAI-food).

Significant and shared stable correlations for component 2 as depicted in figure~\ref{fig:events_comp02} also covered four annotation levels. 
On level 1, the component correlated with four motion categories (13 correlations in total). Its highest number was found for 'release-carry' (6 correlations in CSL-dishes), followed by 'lower' (3 correlations in CSL-dishes) and 'retract-carry' (2 correlations in IAI-dishes) as well as 'grasp-free' (2 correlations in CSL-food).
On level 2, it correlated with the action category of place (3 correlations in CSL dishes). 
On level 3, it correlated with the situation category of deliver (5 correlations in CSL-dishes) 
On level 4, it correlated with four exploratory groups (56 correlations in total). Its highest number found for 'object present' (26 correlations, 19 CSL-dishes, 5 CSL-food, 2 IAI-dishes), followed by 'relation change' (15 correlations, 12 CSL-dishes, 3 CSL-food), 'all arm movement' (8 correlations in CSL-dishes) and 'contact change' (7 correlations, 4 CSL-food, 3 CSL-dishes).

Further significant and stable correlations were found for component 4 in two exploratory groups on annotation level 4 (fig~\ref{fig:events_comp06}; 19 correlations in total). Both were found for navigation with the highest number for 'navigate rotation' (16 correlations in CSL-dishes), followed by 'navigate all' (3 correlations in CSL-dishes).

In addition, correlations for component 6  as depicted in figure~\ref{fig:events_comp06} were found on annotation level 4 for the exploratory group 'contact change' (2 correlations in IAI-dishes). 

\section{Discussion}
Out of the four resulting \SOMA~associated components, brain networks represented by components 1 and 2 had shared and significant stable correlations with ontology events for most participants and event classes. For both, associated event classes on annotation levels 1, 2, and 4 were solely based on arm/hand-based object interaction (obligatory navigation involvement in level 3). Out of the minor components 4 and 6, component 4 had a focus on the navigation event of body rotation, while the neuronal network represented by component 6 showed only a minor number of correlations with motions facilitating hand-object contact changes of annotation level 4. 
These data corroborated the findings of an earlier pilot study by Ahrens2021~\cite{ahrens_neuronal_2021} for most of the depicted brain networks.

On the basis of significant shared stable correlations and environmental and contextual factors, the current results allow for a distinction of two main networks within four domains, with three of these domains rooted in their characteristics of ontological association.

The first domain, task sensitivity, consists of the functional separation between the concepts of fetch and deliver on event levels 1-3. The network of component 1 thereby representing the concept of fetch and ranging from the situation level (lvl 3) down to its most prominent action (lvl 2) and most constituting motions (lvl 1), while the same holds true in case of component 2 for the concept of deliver.

The second domain, event structure bias, consists of the contrasting bias of both networks towards either single events and situations (lvls 1-3) connected to fetching of objects as found in component 1, or a more pronounced activity during a generalized set of event classes (level 4) as found in component 2. 

A third domain, context sensitivity, covers the distinction of findings based on contextual factors. For all components, there was a statistical trend towards correlation to events presented in the context of setting dishes and cutlery as opposed to food and drink items, except for level 4 event grouping of component 1. Out of the two main components, brain activity related to component 2 showed a preference for events recorded in the simpler, less everyday-like environment of the CSL. This was at least in part influenced by the 3:2 ratio of videos recorded in this lab, but further analysis into both factors could prove for a fruitful discussion on this topic.

Underlying all other, a fourth domain of functional characteristics separated networks based on brain activity patterns. Brain activity related to component 1 could be primarily classified into visual perception and object recognition~\cite{weiner_anatomical_2016, malikovic_cytoarchitecture_2016}, while activation patterns of component 2 covered brain areas that are associated with, e.g., (visuo)spatial attention~\cite{ardila_language_2015}, event boundary perception~\cite{zacks_human_2001} as well as action planning and execution, including areas hypothesized to house mirror neurons~\cite{culham_role_2006, maranesi_cortical_2014}.
Additional networks were represented by component 4, which covered areas seen involved in, e.g., whole scene and event boundary perception~\cite{takahashi_pure_2002, zacks_human_2001} as well as mental navigation and episodic memory retrieval~\cite{ghaem_mental_1997, krause_episodic_1999} and component 6 which covered a network involved in primary motor- and somatosensory function and inner verbalization~\cite{shergill_modulation_2002}.
Associated brain networks thereby give potential explanations for correlation characteristics, e.g., prominent focus on the rotation part of navigation events in component 4 due to its network's potential involvement in perception of boundaries between object interaction and navigation.

They also offer avenues for helping understand the characteristics of the other domains. Examples for components 1 and 2 include: 
Fetching-related event classes on levels 1-3 are potentially processed on a more perceptual grade compared to those related to delivery which are more closely associated with the planning- and executive network (task sensitivity domain). The planning- and execution network additionally trends towards generalized event classes while the perception-based network focuses on more specialized classes (event structure bias domain). The planning- and execution focused network might furthermore differ in activity based on the environment of video presentation, and both are more active for specific object or event characteristics that are related to dishes \& cutlery within the context of everyday activities (environmental sensitivity domain).

We now turn to possible interpretations of the results in relation to validating and improving \SOMA. One observation is that the kind of events described by \SOMA~also correlate well to patterns of observed neuronal activity in human subjects, but more fine-grained conclusions can be obtained based on the domains we described above. 
The first domain, task sensitivity, validates a particular task distinction made in \SOMA, i.e., a functional separation of fetch and deliver concepts.

Meanwhile the third domain, context-related biases, suggests some potential additions to \SOMA. While further investigations are required to rule out causes unrelated to brain processing, it appears the brain processes delivery tasks involving dishes differently than delivery tasks involving food. \SOMA~assumes that the roles played by participants in an activity are specifiable depending on the activity itself, e.g., a delivery task defines a 'patient' role played by the object being delivered. While there are different kinds of patient roles, e.g., cut-object which is a patient role defined by cutting tasks, the attribution of a role to an object is not informed by the type of object nor its potential subsequent roles in future activities. This ontological commitment might need rethinking, assuming new data from neuronal investigations strengthen the case for revision.

Another potentially important observation is the trending of certain event classes towards a dominance of being processed by networks focused either on perception or planning and execution.
This distinction becomes apparent when stable correlations during fetching are compared with ones during delivering events. In the present study, fetching tasks have shown significant stable correlations with perceptual networks, while delivery was preferentially correlated with networks that are related to planning and execution.
These results suggest that humans exhibit different neuronal treatment for actions that involve imagined entities such as states which are desired by an agent planning its next action.
This fundamental distinction is only covered sparsely by the \SOMA~ontology.
The task taxonomy of \SOMA~distinguishes between physical and mental tasks where the latter only includes actions whose execution does not involve the agent's body.
The ontology further classifies physical tasks based on goals of the agent.
However, these task types are not grouped according to whether their execution involves mental entities such as an imagined placement of an object.
The results suggest that such a grouping could also be useful in the \SOMA~ontology.
However, further studies will be needed that cover additional action types.

Additional studies would also prove fruitful to gain deeper insights into other characteristics of the current results. For example, only few significant correlations were found with primary motor- and somatosensory networks as represented by component 6, which could have stemmed from the study’s focus on stable and shared correlations. Analyses concerning inter-subject variation or changes in correlation strength due to repetition- and learning effects through repeated video presentations might offer valuable insights into underlying neuronal processing characteristics when compared with other networks.
Furthermore, since component signals were correlated with events defined by our existing ontology, components could be present in the current data-set whose underlying brain networks were stimulus coupled in novel ways not yet covered by existing ontological classes. Analyses focusing, e.g., on effects of resting periods on component signals could help identifying these networks.

Other avenues for extended analysis concern the preferential correlation to events based on contextual factors. 
Underlying stimuli difference could thereby be based in both the spatial dimension, e.g., overall environmental complexity or object characteristics, and/or the temporal dimension, e.g., differences in the timing and distribution of scenes and events. An analysis of the stimulus material concerning these factors would thus prove insightful.
Finally, to achieve \SOMA-related add-on value from the second domain, network bias towards specific vs generalized events, additional knowledge about the temporal nature of both networks is crucial. For example, the planning and execution focused network might exhibit a primarily tonic signal with additional spiking activity during specific object deliverance events for all participants. However, it could also trend towards one or the other for certain groups of participants.

\section{Conclusion}

The aim of the present study was to validate \SOMA~with neuronal activity patterns derived from human volunteers who view recordings of actors who conduct several table setting scenarios. This is, to our knowledge, a fairly new line of research and as such there is significant work still to be done, including at a methodological level -- e.g., does viewing the same video a second time change which networks get activated? If so, how would that bias stable correlations? As such, we have not gathered data to validate the complete \SOMA; however, the early results we have are encouraging. The data derived from ICA and subsequent correlational analyses offer promising insights into domains of human neuronal networks and their relation to \SOMA~underlying our robot cognitive architecture. 
Significant and stable correlations were found with various action and situational levels of the ontological concepts. These correlations were also shared between subjects. Furthermore, a functional distinction in the \SOMA~between the concepts of fetching and delivering were
also represented in the neuronal data, substantiating the ontology’s validity on an organizational level. Additional neuronal network characteristics not yet present in the ontology were found, which are planned to be integrated into the \SOMA~concepts, thus bringing it closer to human neuronal information processing.

\ack The research reported in this paper has been supported by the German Research Foundation DFG, as part of Collaborative Research Center 1320 EASE - \emph{Everyday Activity Science and Engineering}. 
We would like to thank the group of Tanja Schultz for allowing access to her lab facilities for video recordings as well as for providing support with EASELAN. Further, we are grateful to Angela Kondinska for assisting during data acquisition.

\bibliography{literature_id1554}
\end{document}